\documentclass[11pt,a4paper]{article}
\usepackage{amsmath,amssymb}
\usepackage{epsfig,graphicx}
\title{\bf Fluctuation Diffusion in Quark matter}
\author{B.~Kerbikov\footnote{{\bf e-mail}:
borisk@itep.ru}\\
\small{\em State Research Center}\\ \small{\em Institute for
Theoretical and Experimental Physics } \\ \small{\em Moscow,
Russia}\\
{ Talk at the 15th International Seminar "Quarks-2008"}}

 \date{}

\newcommand{\beq}{\begin{eqnarray}}
 \newcommand{\eeq}{\end{eqnarray}}
  \newcommand{\be}{\begin{equation}}
\newcommand{\ee}{\end{equation}}
\def\la{\mathrel{\mathpalette\fun
<}} \def\ga{\mathrel{\mathpalette\fun >}}
\def\fun#1#2{\lower3.6pt\vbox{\baselineskip0pt\lineskip.9pt
\ialign{$\mathsurround=0pt#1\hfil ##\hfil$\crcr#2\crcr\sim\crcr}}}

\newcommand{\vek}{\mbox{\boldmath${\rm k}$}}

\newcommand{\vep}{\mbox{\boldmath${\rm p}$}}

\newcommand{\vegamma}{\mbox{\boldmath${\rm \gamma}$}}

\begin{document}

\maketitle
\begin{abstract}
A unique transition state is a precursor of quark matter
formation. We point out the hallmarks of this  state and evaluate
the quark diffusion  coefficient in the vicinity of the QCD
critical line.
\end{abstract}

During the last decade the investigation of the quark matter at
finite temperature and density became a compelling topic in QCD.
Drawn in the $(T,\mu)$ plane with $\mu$ being the quark chemical
potential the QCD phase diagram \cite{1,2} embodies several
domains with quite different and sometimes poorly understood
properties.  The critical line starts at the point $(T=T_c\simeq
170$ MeV, $\mu=0$) and terminates at $(T=0, \mu\simeq 300-500$
MeV). From $T_c$ with $\mu$ increasing the critical line
presumably corresponds to the analytic crossover which ends at the
critical point of the second order from which the first order
transition line is perceived to originate supposedly ending at $(T=0, \mu\simeq
300-500$ MeV). The  whole critical line spans over the region of
strong coupling QCD regime which fails us for the first principle
calculations. Lattice simulations have been performed along the
$T$ axis at $\mu=0$ while models like NJL have been used to
investigate the transition in the vicinity of the other end point
of the critical line.

Lattice simulations have been recently extended to nonzero but small
values of $\mu, \mu/T\la 1$ \cite{4}. At  the other end of the
critical line at small $T$ and moderate $\mu$ model calculations
suggest that the system is unstable with respect to the  formation
of quark-quark Cooper pair condensate \cite{5,6}. It took quite
some time to realize that the nonzero value of the gap obtained
within the NJL type calculations for $\mu\simeq 300-500$ MeV does
not mean the  onset of color superconducting regime similar to the
BCS one \cite{7,8}.

Nonzero value of the gap is only a signal of the presence of
fermion pairs. Depending on the strength of  the interaction,  on
the fermion density, and on the temperature such pairs may be
either stable, or fluctuating in time, may form a BCS condensate,
or  a dilute Bose gas, or undergo a  Bose-Einstein condensation.
It was shown \cite{8,9} that the critical line at small $T$ and
moderate $\mu$ corresponds to the crossover from strong  coupling
regime of composite nonoverlapping  bosons (diquarks) to the weak
coupling regime of macroscopic overlapping Cooper pair condensate
(with possible LOFF phase \cite{10} in between the two regimes).
The dimensionless crossover  parameter is $n^{1/3}\xi$, where $n$
is the quark density, and $\xi$ is the characteristic length of
pair correlation when the system is in the BCS regime and the root
of the mean-square radius of the bound state when the system is in
the strong coupling regime. The crossover (called BEC-BCS
crossover) occurs at $n^{1/3}\xi \sim 1$. It can be shown
\cite{11} that the same dimensionless parameter defines the
Ginzburg-Levanyuk number $Gi$ which characterize the fluctuation
contribution to the physical quantities and the width of the
fluctuation region

\be Gi=\frac{27\pi^4}{28\zeta(3)} \left(\frac{T_c}{E_F}\right)^4 =
\frac{21\zeta(3)}{64} (k_F\xi)^{-4} \simeq \frac{5\cdot
10^{-2}}{(n^{1/3}\xi)^4},\label{1a}\ee where $\zeta(3) =1.2, T_c
\simeq (0.04-0.05)$ GeV is the critical temperature.
From(\ref{1a}) we obtain $Gi\ga 10^{-2}$, while for the normal
superconductors $Gi\simeq 10^{-12}-10^{-14}$. We see that the
fluctuation effects in a quark system at moderate density are very
strong. The transport coefficients in strong fluctuation regime
can  be expressed in terms of the time-dependent propagator
\cite{12}. In particular, we can derive an analytic expression for
the quark diffusion coefficient. The only relevant parameter
entering into the diffusion coefficient will be the relaxation
time $\tau$. In writing the expression for the fluctuation quark
propagator (FQP) we assume that at moderate values of $\mu$ under
consideration the quark Fermi surface is already formed and hence
momentum integration can be  performed around it in the same way
as in the BCS theory of superconductivity. The FQP is defined as

\be L^{-1}(\vep,\omega) =-\frac{1}{g}+ F(\vep,\omega),\label{1}\ee
\be F(\vep,\omega) = \sum_k G^R (\vek, k_4) G^A(\vep-\vek,
\omega-k_4),\label{2}\ee where  $g$ is the coupling constant with
the dimension $m^{-2}$, the sum over $k$ implies the momentum
integration and Matsubara summation, $k_4=-\pi (2n+1)T, ~G^R$ is
the thermal retarded  Green's function  which reads \be G^R(\vek,
k_4) =i (\vegamma \vek + \gamma_4 k_4 -im + i\mu \gamma_4
-\frac{1}{2\tau})^{-1},\label{3}\ee where $\tau$ is the relaxation
time and $G^A=(G^R)^*$. We shall compute $F(\vep, \omega)$ in the
long-wave fluctuation approximation  \be F(\vep,\omega) \simeq
A(\omega) +B\vep^2,\label{4}\ee where the frequency dependence of
$B$ is neglected.

First we compute \be tr\left\{ G^R (\vek, k_4) G^A(-\vek,
\omega-k_4)\right\}=2\left\{ \frac{1}{\tilde k^2_4+(E-\tilde
\mu)^2} +\frac{1}{\tilde k^2_4+(E+\tilde
\mu)^2}\right\},\label{6}\ee where trace is over the Lorentz
indices, $E^2=\vek^2+m^2,~~ \tilde k_4=k_4-\omega/2,~~ \tilde
\mu=\mu-i\omega/2$.The second term in (\ref{6}) corresponds to
antiquarks and integration around the Fermi surface suppresses its
contribution  though as shown in \cite{11} the interplay of the
quark and antiquark modes may result in instability in the chiral
limit. We shall omit the antiquark contribution and  return to
this question elsewhere. Performing momentum integration around
the Fermi surface we obtain \be A(\omega) =\nu\sum_{n\geq 0}
\frac{1}{\left(n+\frac12+ \frac{\omega}{4\pi
T}\right)},\label{7}\ee where $\mu=2\mu k_F/\pi^2$ is the density
of states at the Fermi surface for two quark flavors. To evaluate
$B$ we act by the operator $(\vep\frac{\partial}{\partial\vek})^2$
on the second Green's function in (\ref{1}). Then we express 
the constant $g$ in (\ref{1}) in terms of
the critical temperature $T_c$. Collecting all the pieces together
we arrive at the FQP containing the diffusion mode \be
L(\vep,\omega) =-\frac{1}{\nu}\frac{1}{\varepsilon +
\frac{\pi}{2T} (-i\omega+ \hat D\vep^2)},\label{8}\ee where
$\varepsilon = \frac{T_c-T}{T_c}$. The diffusion coefficient $\hat
D$ is given by

\be \hat D=-\frac{8T\tau^2v^2_F}{3\pi} \left\{ \psi \left(
\frac12+\frac{1}{4\pi T\tau}\right) -\psi \left(\frac12\right)-
\frac{1}{4\pi T\tau} \psi'
\left(\frac12\right)\right\},\label{9}\ee where
$v_F=\frac{\partial E}{\partial k} (k=k_F)$,  $\psi(z)$ is the
logarithmic derivative of the $\Gamma$-function. The expression is
valid both below and above $T_c$. From (\ref{9}) we obtain the two
limiting regimes
$$ ~~~~~~~~~~~~~~~~~~~~~\hat D\simeq\left\{\begin{array}{llr}  \frac13 v^2_F\tau,& T\tau\ll
1.&~~~~~~~~~~~~~~~~~~~~~~~~~~~~~~~~~~~~~~~~~~~~~~~~~~~~~~~~~~~(10)\\v^2_F/6T,&
T\tau\gg
1.&~~~~~~~~~~~~~~~~~~~~~~~~~~~~~~~~~~~~~~~~~~~~~~~~~~~~~~~~~~~
(11)\end{array}\right.$$


\begin{thebibliography}{99}
\bibitem{1}
M.Stephanov, Progr. Theor. Phys. Suppl. {\bf 153} (2004)
139.

\bibitem{2} R.Casalbuoni, arXiv: hep-ph/0610179.

\bibitem{3} E.Shuryak, arXiv:nucl-th/0609011.

\bibitem{4}O.Philipsen, PoSLAT 2005 (2005) 0.16; arXiv:hep-lat/0510077.

\bibitem{5} K.Rajagopal and F.Wilczek, in " At the Frontier of
Particle Physics", Ed. by  M.Shifman (World Sci., Singapore, 2001)
Vol.3, p.2061.

\bibitem{6} N.Agasian, B.Kerbikov, V.S.Shevchenko, Phys. Rep. {\bf
320} (1999) 131.

\bibitem{7} B.Kerbikov, arXiv: hep-ph/0110197.

\bibitem{8} B.Kerbikov, Phys. Atom.Nucl. {\bf 65} (2002) 1918.

\bibitem{9} B.Kerbikov, Phys. Atom.Nucl. {\bf 68} (2005) 916.

\bibitem{10} A.I.Larkin and Yu.N.Ovchinnikov, ZhETF, {\bf 61} (1971) 2147.
\bibitem{11} B.Kerbikov and E.V.Luschevskaya, Phys. Atom. Nucl.
{\bf 71} (2008) 364.

\bibitem{12} A.Larkin and A.Varlamov, ArXiv: cond-mat/0109177.

\end{thebibliography}
\end{document}